\documentclass[12pt]{iopart}

\usepackage{epsfig}
\usepackage{color}
\usepackage{graphics}
\usepackage{amssymb}
\usepackage{iopams}

\begin{document}

\title[Self-subdiffusion in solutions of star-shaped crowders]{Self-subdiffusion
in solutions of star-shaped crowders: non-monotonic effects of inter-particle
interactions}

\author{Jaeoh Shin $^{\dagger,\ddagger}$, Andrey G. Cherstvy$^\dagger$, and Ralf
Metzler$^{\dagger,\sharp,1}$}
\address{$^\dagger$Institute for Physics \& Astronomy, University of Potsdam,
14476 Potsdam-Golm, Germany\\
$^\ddagger$ Max Planck Institute for the Physics of Complex Systems,
01187 Dresden, Germany\\
$^\sharp$Department of Physics, Tampere University of Technology, 33101
Tampere, Finland}
\ead{$^1$rmetzler@uni-potsdam.de}

\date{\today}

\begin{abstract}
We examine by extensive computer simulations the  self-diffusion of
anisotropic star like particles in  crowded two-dimensional solutions. We
investigate the implications  of the area coverage fraction $\phi$ of the
crowders and  the crowder-crowder adhesion properties on the regime of
transient anomalous diffusion.  We systematically compute the mean squared
displacement (MSD) of the particles, their  time averaged MSD, as well as
the effective diffusion coefficient. The diffusion appears ergodic in the
limit of long traces, such that the time averaged MSD converges towards
the ensemble averaged MSD and features a small residual amplitude spread of
the time averaged MSD from individual trajectories. At intermediate time
scales we quantify the anomalous diffusion in the system.  Also, we show that
the translational---but not rotational---diffusivity of the particles $D$
is a non-monotonic function of the  attraction strength between them. Both
diffusion coefficients decrease  as $D(\phi)\sim (1-\phi/\phi^*)^2$ with
the area fraction $\phi$ occupied by the crowders. Our results might be
applicable to rationalising the experimental  observations of non-Brownian
diffusion for a number of standard  macromolecular crowders used in vitro
to mimic  the cytoplasmic conditions of living cells.
\end{abstract}

\section{Introduction}

Over the recent years, deviations from the standard Brownian diffusion law
\cite{einstein} have been observed in a broad range of systems \cite{bouchaud,
report,pt,sokolov12,franosch,rm-pccp-14,sokolov15}. Depending on the physics of
the system under consideration, various theoretical models are used to describe
these deviations \cite{bouchaud,report,pt,sokolov12,franosch,rm-pccp-14,sokolov15}.
Such \emph{anomalous diffusion\/} is typically characterised by the power-law
growth of the mean squared displacement (MSD) of particles with time
\begin{equation}
\left<\mathbf{r}^2(t)\right>\simeq K_\beta t^\beta.
\end{equation}
We distinguish subdiffusion for $0<\beta<1$ and superdiffusion for $1<\beta$.
Subdiffusion is an abundant phenomenon for passive motion in the world of live
biological cells \cite{pt,sokolov12,franosch,rm-pccp-14,sokolov15}. In the biological
context subdiffusion was observed for particles ranging from small proteins
\cite{fradin05} via messenger RNA molecules \cite{goldingcox} in the cell cytoplasm
to large chromosomal loci and telomeres in the nucleus \cite{telosubdiff} to
sub-micron virus particles \cite{brauch01} as well as lipid granules \cite{jeon11}. 
The features of anomalous diffusion depend on the energy landscape and the
physico-chemical interactions in the system of particles \cite{katja11prl,
katja14sm}. The advances of modern single particle tracking experiments
\cite{goldingcox,lubensky-science,mckintosh-science} provide a wealth of high
resolution experimental data to quantitatively compare the microscopic mechanisms
of non-Brownian diffusion with known theoretical models. The latter include,
inter alia, the continuous time random walk \cite{mw69,sm75,sokol09epl,ad-ctrw-1,
ad-ctrw-2} or the equivalent formulation in terms of fractional diffusion equations
\cite{report,mebakla}, fractional Brownian motion \cite{ad-fbm}, heterogeneous
diffusion processes \cite{hdp}, scaled Brownian motion \cite{rm-sbm,hdp15,usbm}, 
as well as the fractional Langevin equation related to the viscoelasticity of the
environment \cite{visco,visco-weiss}.

The cytoplasm of biological cells is a superdense \cite{goldingcox} fluid
consisting of proteins, nucleic acids, membranous structures, cellular machinery
components, semiflexible filaments, etc. \cite{crowd2,minton08,denton14,weiss14}.
This macromolecular crowding (MMC) reaches volume occupancies of $\phi\gtrsim 30\%$
\cite{elcock-ecoli-cytoplasm}. In addition, the cytoskeletal meshwork
\cite{bausch-new-cyto} of eukaryotic cells impedes the diffusion of larger entities
in cells, in particular, near the cell's plasma membrane. The cytoplasm in addition
is highly heterogeneous both in prokaryotic and eukaryotic cells
\cite{langowski-plos,elf-pnas,cell-fluid}.
The anomalous diffusion of cell-related phenomena may represent a blend of more
than one theoretical model representing the quality of the diffusion on different
length and timescales \cite{pt,sokolov12,rm-pccp-14,weiss09prl,jeon13njp,carlo,
weigel11pnas,tabei13pnas,jeon11prl}.

\begin{figure*}
\includegraphics[width=16cm]{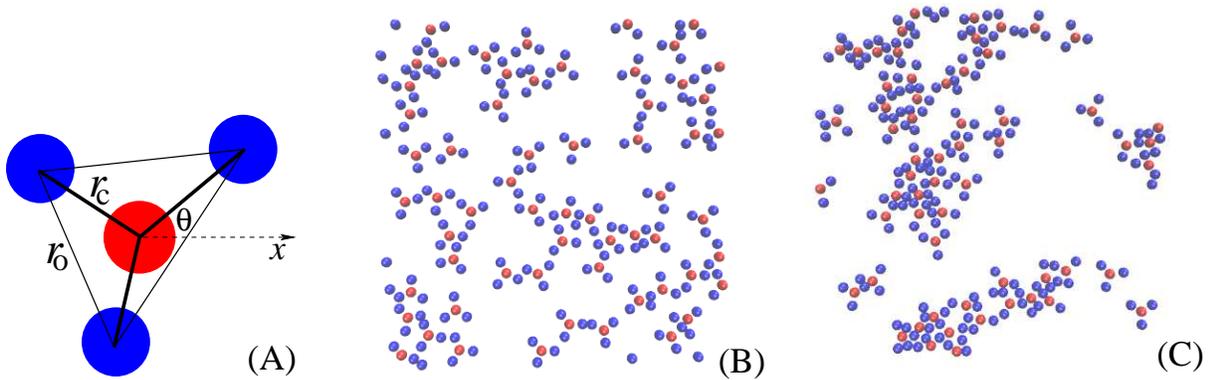}
\caption{(A) Mercedes-Benz$^{\circledR}$ star shaped crowder, with the centre
monomer in red and flexible arms in blue. Typical conformations of crowders for
(B) purely repulsive and (C) attractive interactions of strength $\epsilon_A=1.75
k_BT$ at crowder fraction $\phi=0.12$. Video files illustrating the dynamics of
the stars at $\epsilon_A=0,1,2 k_BT$ are provided in the Supplementary Material.}
\label{fig-config}
\end{figure*}

A number of experimental \cite{langowski-plos,weiss14-pccp}, theoretical
\cite{saxton14}, and simulation \cite{elcock-ecoli-cytoplasm,berry14-pre,
travato-2014,gel-tracer-aljash,szleifer14,shin14njp,shin15sm,shin15acs,thir15a,
thir15b,polymer-dynamics-crowding} studies in recent years were
devoted to tackling various aspects of particle diffusion in crowded
environments. From the simulation perspective, for instance, the studies of
tracer diffusion in non-inert \cite{non-inert-surya}, heterogeneously distributed
and poly-disperse \cite{non-polydisp-surya}, restrictively mobile
\cite{berry14-pre}, squishy \cite{saxton14} and anisotropic
\cite{bagchi-theory,sokolov14jcp} obstacles were performed. Despite the progress
of analytical theories of crowded solutions some important diffusive characteristics
can only be studied quantitatively by computer simulations. This is particularly
true for crowders of the non-trivial of the Mercedes-Benz$^{\circledR}$ star like
particles considered in the current paper (Fig.~\ref{fig-config}). 

We here use computer simulations to unravel the implications of the particle shape
and squishiness as well as the crowding fraction on the
translational ($D$) and rotational ($D_r$) particle diffusivities in highly crowded
solutions. Our main target is to gain insight into the physical behaviour of
non-spherical crowders relevant for the situation in vitro where soft non-spherical
and often non-inert crowders such as globular PEG and branched dextran polymers are
routinely used to mimic the effects of MMC in living cells. Another important
experimental example is the diffusivity of anisotropic lysozyme-like proteins
studied by Brownian Dynamics simulations in crowded media \cite{prot-rotat-polska}.
It was demonstrated that---particularly in heavily crowded solutions---not only a 
transient subdiffusion of the protein centre of mass exists, but diffusion becomes
also progressively anisotropic. This anisotropy of the translational diffusion 
pronounced on short-to-intermediate times disappears in the long time limit.
The long time diffusivity values were shown to drop drastically with the protein
concentration \cite{prot-rotat-polska}. Moreover, the reduction of $D_r$ for
Y-shaped proteins such as IgG $\gamma$-Globulin (molecular weight MW$\approx155$
kDa) was shown to be stronger than for more spherical proteins such as Bovine serum
albumin (MW$\approx$66 kDa). These experimental observations based on fluorescence
correlation spectroscopy measurements are supported by all atom Brownian Dynamics
simulations \cite{prot-rotat-polska}. The inclusion of hydrodynamic interactions
revealed an additional  reduction of $D_r$ of proteins \cite{wade-crowder-shape-BJ}.
The reader is also referred to the simulation study of
Ref.~\cite{winkler-stars-diffusion} in which the self diffusion of star like
polymers in the presence of hydrodynamic interactions \cite{kapral12} was examined
in detail. 

The paper is organised as follows. In section \ref{sec-model} we introduce our
simulation model, the physical observable we are interested in, and some details
on the data analysis algorithms. We present the main findings of simulations in
section \ref{sec-results}.  In section \ref{sec-discussion} the implications of
our results for some cellular systems are discussed.

\begin{figure*}
\includegraphics[width=16cm]{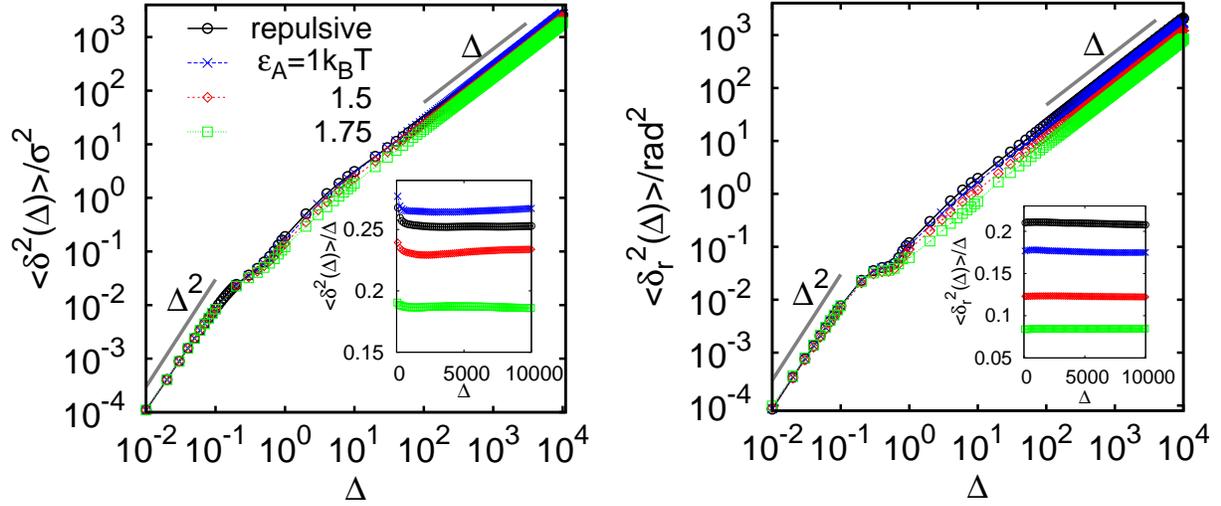}
\caption{Translational and rotational time averaged MSD of star like crowders for
varying  strength of the inter-particle attraction strength $\epsilon_A$. For the
time averaged MSD $\overline{\delta^2(\Delta)}$ only the $x$ components
$\overline{\delta^2_x(\Delta)}$ are shown---the $y$ components show identical
features. The insets show the translational and rotational particle diffusivities
in the long time limit.  Parameters: $\phi=0.15$, $T=2\times10^6$, the average
$\langle\overline{\delta^2_x(\Delta)}\rangle$ is computed over $N=40$ traces.}
\label{fig-msd-f02-eps}\end{figure*}

\section{Simulation model and observables}
\label{sec-model}

We implement our computer code developed to simulate the particle diffusion of
crowded solutions in which all particles are explicitly treated \cite{shin14njp,
shin15sm,shin15acs}. Here, we consider a two-dimensional system of
Mercedes-Benz$^{\circledR}$ star shaped crowders, each consisting of four
discs of diameter $\sigma$ connected by elastic springs, see Fig.~\ref{fig-config}A.
The elastic potential between the midpoint of the molecule and the centres of the
outer monomers is
\begin{equation}
U_c(r)=\frac{1}{2}k_s(r-r_c)^2,
\end{equation}
where $r_c$ is the equilibrium distance and $k_s$ the spring constant. We also
connect the outer monomers with springs of the force constant $k_s$, namely,
\begin{equation}
U_o(r)=\frac{1}{2}k_s(r-r_o)^2, 
\end{equation}
to mimic the softness of our triangular star like crowders. The equilibrium 
distances and constants are set to $r_c=1.5\sigma$, $r_o=1.5 \sqrt{3} \sigma$, 
and $k_{s}=100 k_{B}T/\sigma^2$. The interaction between all beads is described
by the 6-12 Lennard-Jones potential
\begin{equation}
\nonumber
U_{\mathrm{LJ}}(r, r_{\mathrm{cut}})=4\epsilon\left[-\left(\frac{\sigma}{r}\right)^6
+\left(\frac{\sigma}{r}\right)^{12}\right]\Theta(r_{\mathrm{cut}}-r)+C(r_{\mathrm{
cut}}).
\end{equation}
Here $\Theta(x)$ is Heaviside step function and $C(r_{\mathrm{cut}})$ is a constant
setting $U_{\mathrm{LJ}}(r > r_{\mathrm{cut}})=0$. For a purely repulsive potential
the standard cutoff distance $r_{\mathrm{cut}}=2^{1/6}\sigma$ is used with the
potential strength $\epsilon=k_{B}T$. For attractive interactions we set $r_{\mathrm{
cut}}=2\sigma$ with varying adhesion strength $\epsilon=\epsilon_{A}$ between the
monomers. This attraction acts between all the monomers of the stars.
We use periodic boundary conditions within a square box of area $L^2$.
The packing fraction of $N$ crowders in the system is defined as $\phi=N A/{L^2}$,  
where $A=4\pi(\sigma/2)^2$ is the total area of the four monomers and $N\sim10^2$
is a typical number of stars used in our simulations. In most scenarios below the
system size is $L=40\sigma$ and the total simulated trace length is $\sim4\times
10^8$ of elementary time steps.

The dynamics of the two-dimensional position $\mathbf{r}_i(t)$ of the $i$th monomer
disc interacting with the other monomer discs is described by the Langevin equation
\begin{equation}
m\frac{d^2\mathbf{r}_i(t)}{dt^2}=-\gamma\frac{d \mathbf{r}_i(t)}{dt}-
\sum_j \bnabla[U_c(r_{ij})+U_o(r_{ij})+U_{\mathrm{LJ}}(r_{ij})]+\bxi_{i}(t).
\label{eq-langevin}
\end{equation}
Here $\bxi(t)$ represents Gaussian white noise with zero mean $\left<\bxi(t)\right>
=0$ and correlator $\left<\bxi(t)\cdot\bxi(t')\right>=4\gamma k_BT\delta (t-t')$,
where $k_B$ is the Boltzmann constant, $\gamma$ is the friction coefficient, and $T$
the absolute temperature. In the following, we use $\sigma$ and $k_{B}T$ as the
basic units of length and energy, respectively. We simulate the system with the
Verlet velocity algorithm with elementary time step $\Delta t=0.005$ for the total
time $T$. The physical time scale in these simulations is the standard combination
$\delta\tau=\sigma\sqrt{m/(k_B T)}\approx0.36$ ns \cite{simul-book}, if we set the
monomer diameter to $\sigma =6$ nm and its mass to the average mass of cytoplasmatic
crowders, MW$\approx68$ kDa \cite{szleifer14,68crowd}.

We track the positions of the centre monomers of all the crowder stars and their
orientation with respect to the $x$-axis, denoted as $\theta_i$. From the
trajectory of the $i$th crowder we calculate the time averaged translational
($\overline{\delta^2_i}$) and rotational ($\overline{\delta^2}_{r,i}$) MSDs as
\cite{rm-pccp-14}
\begin{eqnarray}
\nonumber
\overline{\delta^2_{i}(\Delta,T)}&=&\frac{1}{T-\Delta}\int_0^{T-\Delta}\{[x_{i}(t+
\Delta)-x_{i}(t)]^2+[y_{i}(t+\Delta)-y_{i}(t)]^2\}dt\\
&=&\overline{\delta^2_{i,x}}+\overline{\delta^2_{i,y}}
\label{eq-tamsd}\end{eqnarray}
and
\begin{eqnarray}
\overline{\delta^2_{r,i}(\Delta,T)}=\frac{1}{T-\Delta}\int_0^{T-\Delta}[\theta_{i}
(t+\Delta)-\theta_{i}(t)]^2dt.
\label{eq-tamsdr}
\end{eqnarray}
Here $\Delta$ is the lag time along the trace. In addition to the individual MSDs
$\overline{\delta^2(\Delta)}$ we compute the corresponding averages over the set
of $N$ individual trajectories, 
\begin{equation}
\left<\overline{\delta^2(\Delta)}\right>=\frac{1}{N}\sum_{i=1}^{N}\overline{\delta
_{i}^2(\Delta)},
\end{equation}
as well as their amplitude spread around this mean value.

The diffusion is called ergodic if the ensemble and time averaged MSDs coincide in
the limit $\Delta/T\to0$ and if the spread of $\overline{\delta^2}$ around the mean
approaches the delta function in this limit \cite{rm-pccp-14}. A more accurate
description of ergodicity can be achieved based on the so-called ergodicity breaking
parameter EB. The latter is defined as the variance of the distribution of the
dimensionless variable $\xi=\overline{\delta^2}/\left<\overline{\delta^2}\right>$,
whose precise behaviour as a function of the lag time and the various model
parameters, however, is beyond the computational scope of the current
investigation.

\section{Results}
\label{sec-results}

\begin{figure}
\begin{center}
\includegraphics[width=10cm]{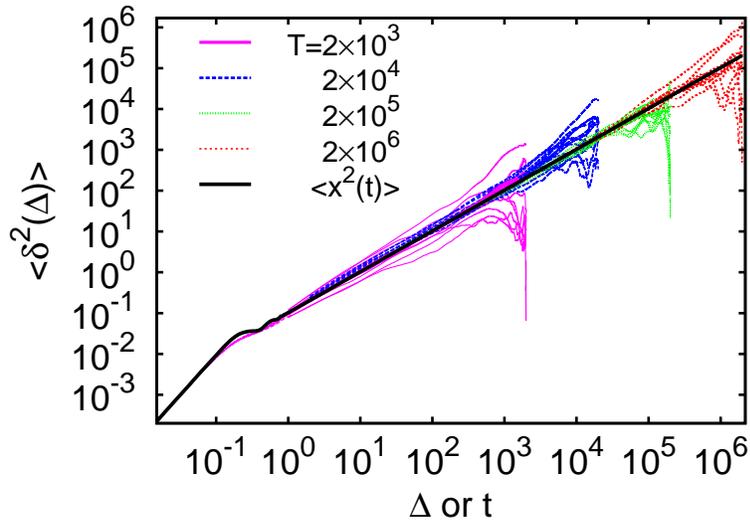}
\end{center}
\caption{Individual time averaged MSD traces and their dependence on the trajectory
length $T$, plotted for the parameters of Fig.~\ref{fig-msd-f02-eps} and $\epsilon_
A=2 k_BT$. The ensemble averaged MSD is the bold black line. The time averaged MSD
trajectories become more reproducible for longer trace length $T$.}
\label{fig-scatter}\end{figure}

\begin{figure}
\begin{center}
\includegraphics[width=8cm]{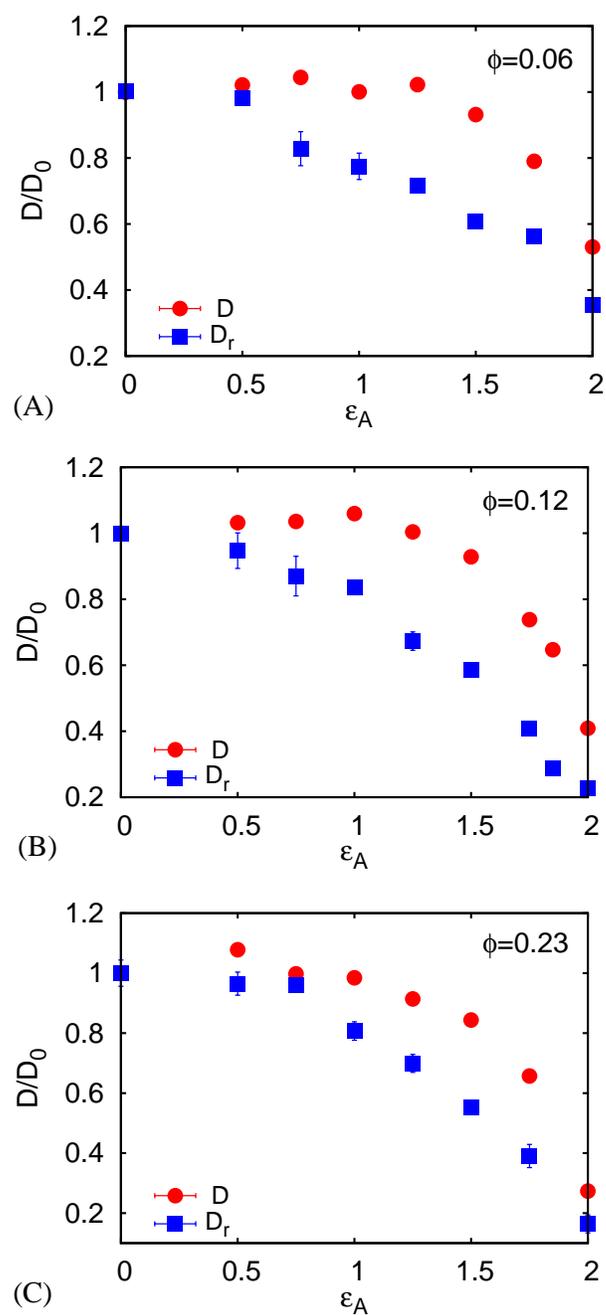}
\end{center}
\caption{Average Brownian diffusivity of crowders measured along the $x$-direction
($D\equiv D_x$ here) versus their mutual attraction strength $\epsilon_A/(k_BT)$,
plotted for the parameters of Fig.~\ref{fig-msd-f02-eps} and varying crowding
fractions $\phi$. The error bars are included and are often smaller than the symbol
size.}
\label{fig-diffusivity-eps}\end{figure}

In Fig.~\ref{fig-msd-f02-eps} we present the behaviour of the translational and 
rotational MSDs for varying inter-particle attraction strength $\epsilon_A$ at
crowder packing fraction $\phi=0.15$. The initial crowder diffusion is ballistic,
stemming from the simulation of inertial particles, see also
Ref.~\cite{non-inert-surya}. At intermediate time scales of $\Delta\sim0.1\ldots10$
we observe a non-monotonic behaviour of the time averaged MSD that we ascribe to the
events of the first collision of a given crowder molecule with another crowder. We
quantify the variation of the local scaling exponent in Fig.~\ref{fig-msd-eps1-f}B
below. In the long lag time limit the translational and rotational MSDs grow
linearly with $\Delta$ reflecting the Brownian behaviour of the crowder particles.
In this limit the diffusion is ergodic, as we demonstrate in Fig.~\ref{fig-scatter}.
This statement is not necessarily trivial: in many weakly non-ergodic systems the
time averaged MSD turns out to be a linear function of the lag time $\Delta$ while
the ensemble averaged MSD scales as a power-law or logarithmically in time $t$. This
phenomenon was observed in various experiments \cite{jeon11,carlo,weigel11pnas,
tabei13pnas}
and explained in terms of various stochastic processes \cite{rm-pccp-14,hdp,rm-sbm,
ctrw,noisy,cctrw,lapeyre,godec-localized,singlefile}. Fig.~\ref{fig-scatter} demonstrates both that to very good
approximation ergodicity in the sense of the equality $\langle x^2(\Delta)\rangle=
\overline{\delta^2(\Delta)}$ and that a very small amplitude scatter around the mean 
$\langle\overline{\delta^2(\Delta)}\rangle$ exists and thus the time averages are
reproducible quantities. We furthermore detail the dependence of the particle
diffusivity in this Brownian limit versus the attraction strength and the filling
fraction in Figs.~\ref{fig-diffusivity-eps} and \ref{fig-diffusivity-f},
respectively. 

Let us be more specific. Fig.~\ref{fig-scatter} illustrates the time averaged MSD
for different lengths of the time series of the star diffusion as well as the
superimposed ensemble averaged MSD shown as the bold black line. As can be seen from
the figure, the amplitude scatter of single traces $\overline{\delta^2}$ around
their mean remains small along the entire trajectory except when $\Delta\sim T$,
as expected. This growing spread as $\Delta\sim T$ is a standard feature of even
canonical Brownian motion appearing due to progressively poorer statistics when
taking the time average \cite{rm-pccp-14}. More importantly we observe that the
amplitude spread of time averaged MSD at a fixed lag time $\Delta$ decreases as
the length $T$ of the time traces increases. This property is ubiquitous for
ergodic diffusion processes \cite{rm-pccp-14}. We note that the magnitude of the
amplitude scatter that we observe for $\overline{\delta^2}$ for moderately adhering
Mercedes-Benz$^{\circledR}$ stars are similar to that of a tracer in a network of
sticky spherical obstacles, compare Fig.~\ref{fig-scatter} above and Fig.~7
as well as the
explanations in the text of Ref.~\cite{non-inert-surya}. Computing the magnitude
of the mean time averaged MSD for varying trace length $T$ we observe that its
magnitude stays nearly unchanged with $T$ (Fig.~\ref{fig-scatter}). In the short
lag time regime the ballistic scaling is visible.
Given these
observations our process is ergodic and thus fundamentally different from other
anomalously diffusive systems such as those described by continuous time random
walks or heterogeneous diffusion processes \cite{hdp}. For the
latter a pronounced scatter of the time averaged MSD trajectories around their
mean and a clear dependence of the amplitude of $\left<\overline{\delta^2(\Delta)}
\right>$ on $T$ at fixed $\Delta$ exist, that is, the system ages \cite{rm-pccp-14,
johannes,hdp_age}. 

\begin{figure}
\begin{center}
\includegraphics[width=10cm]{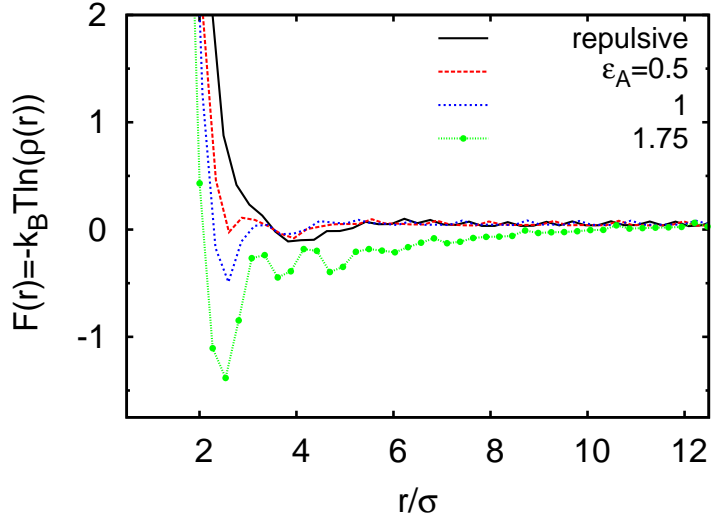}
\end{center}
\caption{Effective potential (\ref{eq-fe}) between the crowders for different
attraction strengths $\epsilon_A/(k_BT)$, plotted for the parameters of
Fig.~\ref{fig-msd-f02-eps} using the same colouring scheme.}
\label{fig-effective-potential}
\end{figure}

The particle diffusivities are defined as
\begin{equation}
D_x=\frac{\left<\overline{\delta^2_x(\Delta)}\right>}{2\Delta}
\end{equation} 
and
\begin{equation}
D_r=\frac{\left<\overline{\delta^2_r(\Delta)}\right>}{2\Delta}
\end{equation}
obtained in the long time limit $\Delta\gg1$. Fig.~\ref{fig-diffusivity-eps} shows
the values of $D$ and $D_r$ obtained from a linear fit of the translational and
rotational time averaged MSDs in the range $\Delta=10^3\ldots10^4$. We find that
while the rotational diffusivity $D_r$ decreases monotonically, the translational
diffusivity remarkably exhibits a shallow yet significant maximum at $\epsilon_A^*
\approx1k_BT$. This systematic trend persists for the variation of the crowder
fraction in a quite broad range (Fig.~\ref{fig-diffusivity-eps}). This implies that
the self-diffusion of our star like crowders can be facilitated by a weak
inter-particle attraction. This is one of the main conclusions of this study. One
can rationalise this trend in the self-diffusion
in terms of the concept of the effective crowder size that decreases for moderate  
attraction strengths $\epsilon_A\approx1k_BT$. Fig.~\ref{fig-diffusivity-eps}
illustrates that for progressively stronger star-star attraction their mutual 
diffusivity decreases eventually to zero due to aggregate formation, see also
Fig.~\ref{fig-effective-potential} and its discussion below.

We also detect a progressive aggregation of crowders at relatively large values of
the crowder-crowder attraction strength $\epsilon_A$, as demonstrated in
Fig.~\ref{fig-config}C and in the video files in the Supplementary Material. This
is a well known phenomenon, for instance, in the glass transitions of dense
suspensions of sticky hard spheres \cite{poon}. Accordingly, the decrease of the 
average diffusivity $D$ of crowders as plotted in Fig.~\ref{fig-diffusivity-eps} 
emerges due to the averaging over an ensemble of particles that perform individual
random motions. This average takes into account both particles forming transient
aggregates as well as free particles. Roughly speaking the average diffusivity drops
inversely proportionally to the number of particles in the cluster. The fraction of
particles clustering in these aggregates increases with the mutual attraction
strength. The average diffusivity therefore progressively decreases with $\epsilon
_A$ due to a larger fraction of particles in the transient aggregates.
At large attraction strength the majority of particles belong to big clusters.

At a fixed cohesiveness $\epsilon_A$ of our Mercedes-Benz$^{\circledR}$ stars
vicinal crowders create a rough energy landscape for the self-diffusion and the
hopping of a given crowder particle. As the MMC fraction $\phi$ increases, the
binding events give rise to more a prolonged particle aggregation and reduced
self-diffusivity. Above a critical MMC fraction the barrier height exceeds the
thermal energy unit $k_BT$ thus increasing the lifetime of crowder aggregates
significantly. For stronger star-star attraction the formation of essentially
permanent aggregates sets in for less crowded systems leading to an inhomogeneous,
phase-separated spatial distribution, see Fig.~\ref{fig-config}C.

A similar non-monotonicity of the translational diffusivity $D$ at similar strengths
of the  particle-crowder attraction was found in Ref.~\cite{szleifer14} for the
tracer diffusion in dense suspensions of spherical Brownian particles. While we
here detect that the attraction strength yielding the highest value of $D$ is a
function of the crowding fraction $\phi$ of the stars for the spherical particles,
the stickiness facilitating the particle diffusivity was almost $\phi$-independent
in Ref.~\cite{szleifer14}. The non-monotonic $D(\epsilon_A)$ dependence was 
interpreted in Ref.~\cite{szleifer14} in terms of the roughness of the free energy
landscape for the tracer diffusion using the concept of the chemical potential. 
Interestingly, the tracer diffusivity was also non-monotonic in $\phi$ in a static 
regular array of sticky obstacles, as quantified in Ref.~\cite{non-inert-surya}.

We checked the universality of the observed dependencies for $D(\epsilon_A)$ and
$D(\phi)$ also for spherical particles. Namely, we simulated just a single monomer
of our Mercedes-Benz$^{\circledR}$ stars with the given adhesive properties. The
diffusivity was indeed found to reveal a maximum at $\epsilon_A^{opt}\sim0.5\ldots
1k_BT$ (not shown), indicating some universality of this a priori counter-intuitive
faster diffusion for a weak inter-particle attraction \cite{szleifer14}. Note also
that for a polymer chain diffusing in an array of sticky obstacles a weak
chain-obstacle attraction can also substantially enhance the polymer diffusivity
\cite{szleifer14,lee02}.

To rationalise the observed behaviour of $D(\epsilon_A)$ we calculate in
Fig.~\ref{fig-effective-potential} the potential of the mean force between two
crowders as 
\begin{equation}
F(r)=-k_BT\log[\rho(r)].
\label{eq-fe}
\end{equation}
In this reconstructed free energy $\rho(r)$ is the average radial distribution
function of the centre monomer of the crowders in the steady state long time limit.
As the mutual attraction strength $\epsilon_A$ increases we observe that the
potential well at the separation $r\approx\sigma$ becomes deeper, see the first
well in Fig.~\ref{fig-effective-potential}. Concurrently, the distance at which
$F(r)$ sharply increases becomes shorter for larger $\epsilon_A$. For a stronger
star-star attraction the crowders feature a more organised appearance, resulting
in measurable oscillations of $\rho(r)$ and $F(r)$, as evidenced in
Fig.~\ref{fig-effective-potential}. These trends indicate that the effective crowder
radius gets smaller with increasing $\epsilon_A$, and at an optimal value $\epsilon
_A^{opt}$ the crowders approach one another more closely yet without sticking. This
in turn might result in a faster average diffusivity $D$ at $\epsilon_A\approx
\epsilon_A^{opt}$, as we indeed observe. An effective reduction of the crowder
size at optimal attraction strength is one important cause---albeit possibly not the
only one---for this facilitated diffusion. In the current system, the equilibrium
distance of the outer monomers from the cenral monomer is reduced by about 2\% for
the inter-monomer attraction strength of $2k_BT$.
Even higher values of $\epsilon_A$ give
rise to the formation of large clusters of crowders, see the Supplementary Material.
As shown in Fig.~\ref{fig-diffusivity-eps} the corresponding diffusivity of an
average particle drops dramatically.

\begin{figure*}
(A)~~\includegraphics[width=14.8cm]{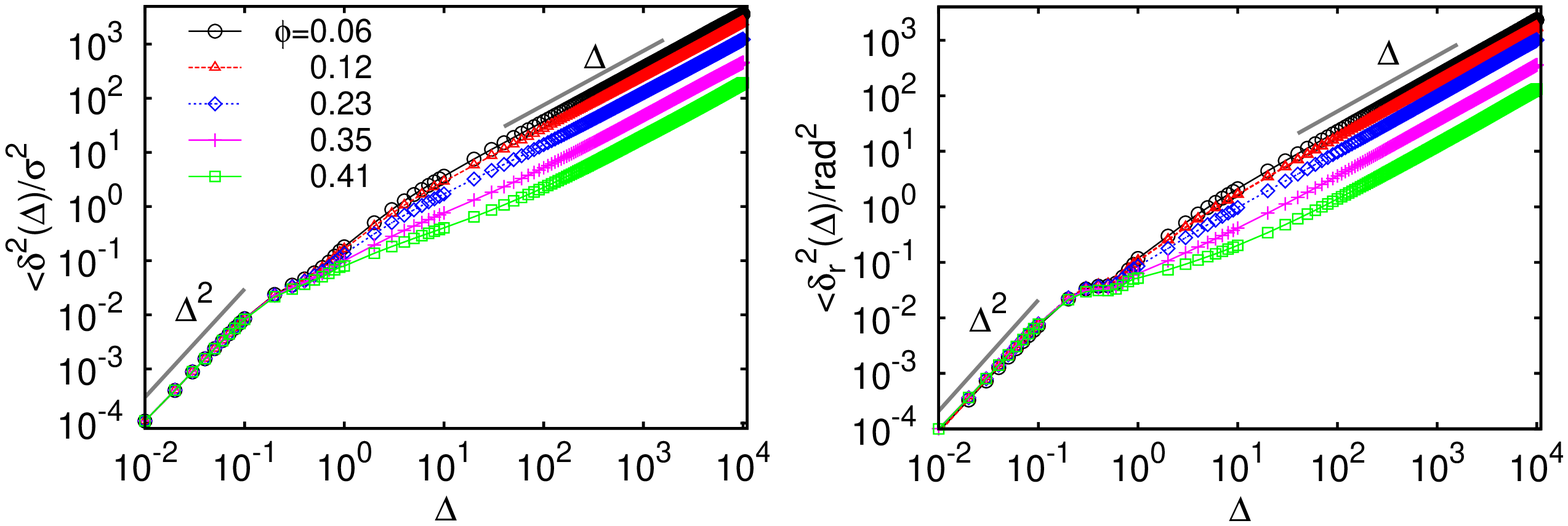}\\
(B)~~\includegraphics[width=14.8cm]{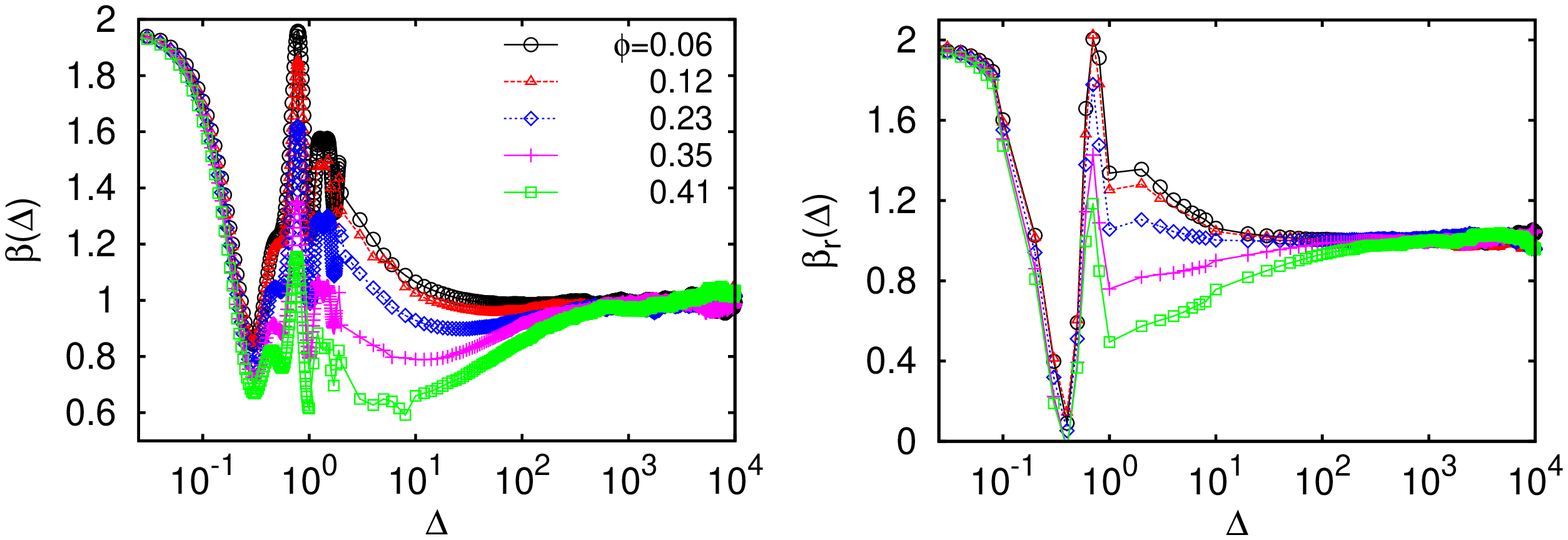}
\caption{A: Translational time averaged MSD of the central star monomer and
rotational time averaged MSD of the star polymer. B: local scaling exponent
$\beta(t)$ of the time averaged MSDs computed for varying packing fractions 
$\phi$. In B, in the limit of short times a linear sampling of data points
was chosen for the left panel and a logarithmic sampling for the right panel.  
Parameters: $\epsilon_A=1k_BT$, $T=2\times10^6$, and $N=10$.}
\label{fig-msd-eps1-f}
\end{figure*}

In Fig.~\ref{fig-msd-eps1-f}A we show the translational and rotational MSD for 
varying packing fraction of crowders  $\phi$. As expected---from a linear fit to
the long time time averaged MSD---the diffusivity is a monotonically decreasing
function of $\phi$, as evidenced by Fig.~\ref{fig-diffusivity-f}. For more crowded
systems the tracer diffusion gets more obstructed and the magnitude of the
corresponding mean time averaged MSD $\left<\overline{\delta^2}\right>$ decreases. 
To elucidate these effects further we evaluate from the time averaged MSD traces
of Fig.~\ref{fig-msd-eps1-f}A for the translational motion the local diffusion
exponent \cite{rm-pccp-14,dagdug-alfa} 
\begin{equation}
\beta(t)=\frac{d\log\left(\left<\overline{\delta^2(\Delta)}\right>\right)}{d
\log(t)}.
\label{eq-power}
\end{equation} 
For the rotational motion the exponent $\beta_r(t)$ is defined analogously. We
observe a ballistic regime with $\beta=2$ in the particle diffusion at short times, 
(Fig.~\ref{fig-msd-eps1-f}B). This ballistic regime is followed by a
decrease and further increase of the scaling exponent at $\Delta\sim1$. 
These non-monotonic trends are also clearly visible from the behaviour of the time 
averaged MSD traces themselves as a function of the lag time $\Delta$, see
Fig.~\ref{fig-msd-eps1-f}A. We find that the variations of the scaling exponent for
translational and rotational motions of the star like crowders appear correlated,
indicating a coupling of these diffusion modes \cite{lubensky-science}. In the
plots for the scaling exponent $\beta(t)$ in Fig.~\ref{fig-msd-eps1-f}B the
significant spike-like signal at $\Delta\sim1$ is interpreted as an effect of the
first collision of particles and the resulting onset of an effective confinement.
We note that even in effective one-particle theories pronounced oscillations occur
at the crossover point between initial ballistic and the overdamped regime
\cite{stas_osc,kursawe}.

With increasing crowder fraction $\phi$ we also observe a more pronounced range of
anomalous diffusion for lag times of the order of $\Delta\sim1\ldots100$. This
range appears strongly correlated between rotational and translational particle
motion, as shown in Fig.~\ref{fig-msd-eps1-f}B. For rotational diffusion the scaling
exponent drops practically to zero for times longer than those of the initial
ballistic growth, and the corresponding mean time averaged MSD trace $\left<
\overline{\delta^2_r}\right>$ exhibits a short plateau (Fig.~\ref{fig-msd-eps1-f}B).
In the long lag time limit the exponent becomes Brownian $\beta\approx1$. Such a
transient subdiffusion was observed for a number of systems \cite{sokolov12,franosch,
rm-pccp-14}, see also the Introduction. Especially in dense colloidal
systems close to the glass transition $\phi=\phi^*$ this subdiffusion is accompanied
by an exponential growth of the solution viscosity $\eta=\eta(\phi)$ which is
divergent at $\phi\to\phi^*$ \cite{glass1}. The colloidal glasses also exhibit
progressive particle localisation effects as discussed in Refs.~\cite{glass1,glass2}.

\begin{figure}
\begin{center}
\includegraphics[width=10cm]{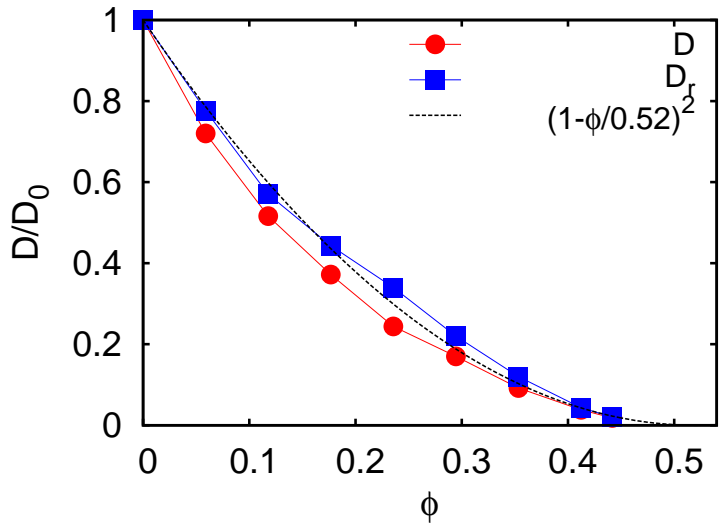}
\end{center}
\caption{Normalised translational (measured along the $x$-direction) and rotational 
diffusivity as function of the crowding fraction $\phi$ computed for $\epsilon_A=1
k_BT$, together with the asymptote of Eq.~(\ref{eq-theory}) with $D(\phi=0)=D_0$.}
\label{fig-diffusivity-f}
\end{figure}

Remarkably, the relative variation of the translational and rotational diffusivities
with the crowding fraction of stars is quite similar. For comparison, we plot in
Fig.~\ref{fig-diffusivity-f} the theoretical prediction for dense suspensions 
of hard spheres \cite{diffusivity94,diffusivity95} 
\begin{equation}
\frac{D(\phi)}{D(0)}=\left[1-\frac{\phi}{\phi^*}\right]^2
\label{eq-theory}
\end{equation}
with the critical packing fraction for our system of $\phi^*(\epsilon_A=1k_BT)
\approx0.52$. Above this value $\phi^*$ both translational and rotational
diffusivities of the crowders essentially vanish. At this critical crowding
fraction the inter-particle attraction becomes so strong that the self-diffusion
is almost completely localised and the motion of particles corresponds more to a
very restricted wiggling and jiggling. As expected, when the star-star interactions
become stronger aggregate formation sets in for less crowded systems, and thus the
critical value $\phi^*$ is diminished (not shown). Albeit this theory in
Refs.~\cite{diffusivity94,diffusivity95} is developed for three dimensional
suspensions in the presence of hydrodynamic interactions, it agrees remarkably well
with our results, as shown in Fig.~\ref{fig-diffusivity-f}. The reader is also
referred to Ref.~\cite{dhont92} for experimental data of the crowding dependent
diffusivity of colloidal particles and alternative theoretical predictions for the
diffusivity $D(\phi)$. We note that Ref.~\cite{streched-expon-diff-crowding}
suggest exponential rather than power-law forms for the particle diffusivity in
crowded solutions.

\section{Discussion and conclusions}
\label{sec-discussion}

We performed extensive computer simulations and theoretical data analysis of the
diffusion of crowders with a branched structure. A simple example of such spiky but
responsive crowders in two dimensions are deformable Mercedes-Benz$^{\circledR}$
like stars employed here. Their outer monomers are inter-connected by an elastic
potential bestowing upon it a certain degree of responsiveness---an important
characteristics for many polymeric crowders \cite{saxton14}. We also incorporated
in the simulations an inter-particle attraction strength which represents another
realistic feature of solutions of non-ideal crowders in vitro. 

We found that the diffusion of our Mercedes-Benz$^{\circledR}$ star like crowders
is ergodic and, within accuracy, Brownian in the long time limit. We examined the
behaviour of the ensemble averaged MSD and the time averaged MSDs of the crowders
in a wide range of the crowder fraction $\phi$ and the inter-crowder attraction
strength $\epsilon_A$. As a function of the crowding fraction we demonstrated that
both translational $D$ and rotational $D_r$ diffusivities follow the analytical 
decrease (\ref{eq-theory}) of $D(\phi)$ predicted for suspensions of hard spheres.
The dependence of the star-star attraction strength is more remarkable. Namely, the
translational diffusivity shows a weak yet systematic non-monotonic dependence on
$\epsilon_A$ for the solutions at all crowding fractions studied herein. The
rotational diffusivity, in contrast, is a monotonically decreasing function of the
inter-particle attraction strength $\epsilon_A$. Thus, a relatively weak
inter-monomer attraction can facilitate the lateral diffusion and also induce a
certain degree of clustering and spatial heterogeneities in crowded solutions of
non-inert particles. These effects will impact the diffusion of a tracer particle
in crowded solutions---such as those of PEG, dextran, or Ficoll---used in vitro to
mimic the crowded conditions in living cells \cite{saxton14,travato-2014}. In
addition to the proof of the ergodic long lag time diffusion shown in
Fig.~\ref{fig-scatter} and the transient subdiffusion regime of our star like
crowders in Fig.~\ref{fig-msd-eps1-f}, Figs.~\ref{fig-diffusivity-eps} and
\ref{fig-diffusivity-f}  for the dependencies of the diffusivities are the
principal results of the current study.

Of course, our planar triangle-like stars still represent a quite primitive system 
to mimic the non-ideal shape of real crowders in experimentally relevant setups. 
Future investigations including a three dimensional pyramid-like shape of crowders
with longer polymeric arms will further elucidate the physical consequences of
non-spherical and squishy crowders, and potentially exhibit additional unexpected
behaviour. Moreover, not only the self-diffusion is to be studied but also the
diffusion of tracer particles of various sizes and shapes in such crowded
suspensions \cite{thir15b} as well as poly-disperse mixtures of crowders
should be investigated. Some asymmetry may also be
incorporated in the crowder shape. Recently, single particle tracking measurements 
allowed one to rationalise the translational and rotational diffusivities of
micron-size symmetric and asymmetric boomerang-shaped particles in two dimensions
\cite{jpcb-recent}. It was observed that the regimes of Brownian
diffusion exist at short and long times while a coupling of $D$ and $D_r$ gave
rise to subdiffusion at intermediate times.

\ack
We acknowledge funding from the Academy of Finland (Suomen Akatemia, Finland
Distinguished Professorship to RM), Deutsche Forschungsgemeinschaft (DFG Grant
to AGC), and the Federal Ministry of Education and Research (BMBF
Project to JS).

\end{document}